\input phyzzx

\def\IR{{\hbox{{\rm I}\kern-.2em\hbox{\rm R}}}}
\def\IB{{\hbox{{\rm I}\kern-.2em\hbox{\rm B}}}}
\def\IN{{\hbox{{\rm I}\kern-.2em\hbox{\rm N}}}}
\def\IC{{\ \hbox{{\rm I}\kern-.6em\hbox{\bf C}}}}

\def\IZ{{\hbox{{\rm Z}\kern-.4em\hbox{\rm Z}}}}
\def\to{\rightarrow}

\def\underarrow#1{\vbox{\ialign{##\crcr$\hfil\displaystyle
{#1}\hfil$\crcr\noalign{\kern1pt
\nointerlineskip}$\longrightarrow$\crcr}}}
%

\def\ltorder{\mathrel{\raise.3ex\hbox{$<$}\mkern-14mu
             \lower0.6ex\hbox{$\sim$}}}
\def\lesssim{\mathrel{\raise.3ex\hbox{$<$}\mkern-14mu
             \lower0.6ex\hbox{$\sim$}}}


\font\ilis=cmbx10 scaled \magstep2
\font\Hugo=cmbx10 scaled \magstep3

\input phyzzx
\overfullrule=0pt
\tolerance=5000
\overfullrule=0pt
\twelvepoint

\twelvepoint

\titlepage
\vskip 2truecm

\centerline{\Hugo On the Weyl-Wigner-Moyal Description} \centerline{\Hugo
of SU$(\infty)$ Nahm Equations} \vglue-.5in \author{ Hugo
Garc\'{\i}a-Compe\'an\foot{ Present address: {\it School of Natural
Sciences, Institute for Advanced Study, Olden Lane, Princeton NJ 08540,
USA.} E-mail: compean@sns.ias.edu} and
 Jerzy F. Pleba\'nski \foot{E-mail: pleban@fis.cinvestav.mx}}
\medskip

\address{Departamento de F\'{\i}sica
\break  Centro de Investigaci\'on y de
Estudios Avanzados del IPN.
~\break Apdo. Postal 14-740, 07000, M\'exico D.F., M\'exico.}

\bigskip

\abstract{ We show how the reduced Self-dual Yang-Mills theory described
by the Nahm equations can be carried over to the Weyl-Wigner-Moyal formalism
employed recently in Self-dual gravity. Evidence of the existence of
correspondence between BPS magnetic monopoles and space-time
hyper-K\"ahler metrics is given.}

\vskip 1truecm

\endpage
\chapter{\bf Introduction}

BPS magnetic monopoles have attracted a great deal of interest in the last
two years. The surprising realization of electromagnetic duality in some
effective $n=2$ supersymmetric Yang-Mills theories in four dimensions opens
various windows for the understanding the non-perturbative aspects of
quantum field theories of point particles and strings [1]. BPS monopoles
are solutions of Bogomolny equations and are particular solutions to the
static, finite energy Yang-Mills-Higgs equations. Many authors working on
the ``mathematical side'' have shown the correspondence of BPS monopoles with
algebraic geometry and twistor theory [2].  Moreover, Nahm has codified
these monopole solutions in terms of a modified version of
Atiyah-Drinfeld-Hitchin-Manin (ADHM) construction [3]. There is a one to
one correspondence between solutions of Bogomolny equations with
appropriate asymptotic boundary conditions and solutions of SU$(2)$ Nahm
equations, $ { d T_i \over ds} = {1 \over 2} \epsilon_{ijk}[T_j,T_k], $
where summation over repeated indices is assumed [2]; (the matrices $T_i$
must, of course, satisfy some consistence conditions). These mathematical
constructions have been recently interpreted in terms of Dirichlet
(D)-3-branes realization of $D=4$ $n=2$ super Yang-Mills theory in type
IIB superstrings [4].  Application of D-instantons to ADHM construction in
terms of monads was presented in Ref. [5].  Furthermore the SL$(2,{\bf
Z})$ duality symmetry of type IIB superstrings was shown to be very useful
tool to study some moduli spaces of BPS monopoles and to explain, among
another subjects, the origin of the``mirror symmetry'' in three dimensions
[6].

On the other hand, Nahm equations also appears in the Ashtekar's approach
to $3+1$ formulation of complex Einstein's equations.  Moreover the
self-dual metrics are in correspondence with solutions to Nahm equations
for a triad of Hamiltonian vector fields [7]. It is very well known that
Moduli space of BPS-saturated configurations with magnetic charge $k$,
${\cal M}_k$, has hyper-K\"ahler structure [2]. Nevertheless not apparent
correspondence exists with the space-time hyper-K\"ahler geometry. The
natural question arises: Is it possible to relate a BPS monopole with a
space-time hyper-K\"ahler four dimensional metric? In this paper we show
that the that answer to this question, in some cases, should be yes. In
order to do that we will employed the Weyl-Wigner-Moyal (WWM) formalism of
quantum mechanics. This formalism has been recently applied to the
self-dual Yang-Mills and the principal chiral model approaches to
self-dual gravity [8,9,10]. In particular, in the last reference of [10],
some explicit solutions were obtained. Then it has been found in [11], how
some general properties of two-dimensional classical chiral models can be
carried over to self-dual gravity. Furthermore, Uhlenbeck's unitons, [12]
which are solutions to chiral equations (with appropriate boundary
conditions), are found to be connected with hyper-K\"ahler metrics,
defining the ``gravitational uniton''.

The study of some large-$N$ limits of various physical systems shows that
radical simplifications occur in their mathematical structures. Some
two-dimensional integrable systems seem to be greatly affected in this
limit, turning out `more integrable'.  Example of this `induced
integrability' occurs in the large-$N$ limit of SU$(N)$ Nahm equations
[13]. Following Witten [14], one would imitate Ward's procedure of
SU$(\infty)$ Nahm equation results, to get integrability, for instance, in
full Yang-Mills theory. Here is where WWM-formalism might be of some
interest in order to address this problem.

Besides of self-dual gravity we find in large-$N$ limit of BPS monopole
equations another example where the WWM-formalism can be applied.

The paper is organized as follows. In Section 2 we review the large-$N$
limit of Nahm equations given in Refs. [13,15]. Section 3 is devoted to
find the Moyal deformation of Nahm equations using the WWM-formalism.  In
Section 4 we find some solutions of the Moyal deformation of Nahm
equations and for the SU$(\infty)$ case as well.  Also we comment about
the possible relation between BPS monopoles and particular hyper-K\"ahler
metrics. Finally in Section 5 we give our final remarks.

\vskip 1truecm

\chapter{\bf Preliminary of Nahm Equations}

\section{  SU$(N)$ Nahm Equations}

In this section we briefly describe the geodesics on SU$(N)$.  Basically
we follows Ward's papers [13] and the finite dimensional group version of
the description of self-dual membranes in $(4+)1$-dimensions by using
self-dual Yang-mills theory on the four-dimensional Euclidean space-time
[15].

Let ${\bf G}$ be an $d$-dimensional Lie group and ${\cal G}$ its
corresponding Lie algebra. Let $A^j(t) \in {\rm so}(3) \otimes {\cal G}$
be a vector field on ${\bf G}$ and $j$ is a so$(3)$ index. In the temporal
gauge connection, $A_0=0$, full Yang-Mills equations are reduced to a
system with Lagrangian [15]

$$ L_N= \alpha {\rm Tr}\bigg\{ A^j_{,t} A^j_{,t} + {1 \over 2} [A^j,A^k]
[A^j,A^k]\bigg\}, \eqno(1)$$ where $A^j_{,t} \equiv {\partial A^j \over
\partial t}$, etc., $\alpha$ is a constant and `Tr' stands for a bilinear
invariant form on the Lie algebra ${\cal G}$ which we will assume to be
semi-simple, and the constraint

$$ [A^j_{,t},A^j] = 0, \eqno(2)$$
which is the Gauss law.

The equations of motion of (1) are

$$ A^j_{,tt} + [A^k,[A^k,A^j]] = 0. \eqno(3)$$
>From the Lagrangian (1) the `conserved energy' of this system is

$$ E_N = {\rm Tr}\bigg\{  A^j_{,t} A^j_{,t} - {1 \over 2}[A^j,A^k]
[A^j,A^k]
\bigg\}, \eqno(4) $$
which is not positive-definite.

On the other hand it is easy to see that  Eqs. (2) and (3) are
implied by the reduced version of self-dual Yang-Mills equations

$$ A^j_{,t} = {1 \over 2} \epsilon^{jkl}[A^k,A^l]. \eqno(5) $$
These equations are the famous Nahm equations and their solutions imply
vanishing of energy in (4).

\section { SU$(\infty)$ Nahm Equations}

The transition from SU$(N)$ to the SU$(\infty)$ gauge theory involves the
substitution, in the place of the Lie algebra su$(N)$, the area-preserving
diffeomorphisms algebra (or Poisson bracket algebra) sdiff$(\Sigma)$, on
a two-dimensional manifold $(\Sigma, \omega)$ [16,17,18]. Here $\Sigma$ is
a two-dimensional simply connected symplectic manifold with local real
coordinates $\{p,q\}$. Locally, by the Darboux's theorem, the symplectic
structure is given by $\omega = dp \wedge dq$.  sdiff$(\Sigma)$ is
precisely the Lie algebra associated with the infinite dimensional Lie
group, SDiff$(\Sigma)$, which is the group of diffeomorphisms on $\Sigma$
preserving the symplectic structure.

The Hamiltonian vector fields are ${\cal U}_{{\cal H}_i} =
~\omega^{-1}(d{\cal H}_i)$ satisfying the sdiff$(\Sigma)$ algebra $[{\cal
U}_{{\cal H}_i},{\cal U}_{{\cal H}_j}] = {\cal U}_{\{{\cal H}_i,{\cal
H}_j\}_P}, \ \ \ {\rm for \ all } \ (i \not = j)~$. Here $\{, \}_P$ stands
for the Poisson bracket. Locally it can be written as $\{{\cal H}_i,{\cal
H}_j\}_P = \omega^{-1}(d{\cal H}_i,d{\cal H}_j) = \omega^{lm} \partial_l
{\cal H}_i \partial_m {\cal H}_j,$ where $\partial_l \equiv {\partial
\over \partial x^l}$, $(l=0,1),$ $x^0 = p,$ $x^1 = q$ and ${\cal H}_i=
{\cal H}_i(p,q).$ The generators of sdiff$(\Sigma)$ are the Hamiltonian
vector fields $ {\cal U}_{{\cal H}_i} = {\partial {\cal H}_i \over
\partial q} {\partial \over \partial p} - {\partial {\cal H}_i \over
\partial p} {\partial \over \partial q},$ associated to the Hamiltonian
functions ${\cal H}_i$.

Considering now the case when $N \to \infty$.  Then the corresponding Lie
algebra ${\cal G} = {\rm su}(\infty)$ should be replaced by the Poisson
bracket algebra on $\Sigma$ [19]

$$ {\rm su}(\infty) \cong  {\rm sdiff}(\Sigma). \eqno(6)$$

Then one can take  $A^j$ to be a triad of Hamiltonian fields

$$ A^j = {\partial {\cal A}^j \over \partial q} {\partial \over \partial p} -
{\partial {\cal A}^j \over \partial p}{\partial \over \partial q},
\eqno(7)$$
where ${\cal A}^j = {\cal A}^j(t,p,q) \in {\rm so}(3)\otimes {\rm
sdiff}(\Sigma).$

Now following the standard prescriptions in the large-$N$ limit in gauge
theories

$${ (2 \pi)^4 \over N^3} {\rm Tr}(...) \Rightarrow - \int_{\Sigma}(...)
dp dq, $$ $$ A^j \Rightarrow {\cal A}^j, \eqno(8)$$ $$ [A^j,A^k]
\Rightarrow \{{\cal A}^j,{\cal A}^k \}_P $$ and taking $ \alpha = { (2
\pi)^4 \over N^3} $, the large-$N$ limit equation of motion (3) is

$$ {\cal A}^j_{,tt} + \{{\cal A}^k,\{{\cal A}^k,{\cal A}^j\}_P\}_P = 0,
\eqno(9) $$
and the constraint (2) reads

$$\{{\cal A}^j_{,t},{\cal A}^j\}_P = 0. \eqno(10) $$

Eqs. (9) and (10) can be derived from the Lagrangian

$$ L^{\infty}_N= - \int_{\Sigma} dp dq \bigg\{ {\cal A}^j_{,t} {\cal
A}^j_{,t} + {1 \over 2} \{{\cal A}^j,{\cal A}^k\}_P \{{\cal A}^j,{\cal
A}^k\}_P \bigg\}. \eqno(11) $$

The large-$N$ limit of `conserved energy' is $$ E^{\infty}_N = -
\int_{\Sigma} dp dq \bigg\{ {\cal A}^j_{,t} {\cal A}^j_{,t} - {1 \over
2} \{ {\cal A}^j,{\cal A}^k \}_P \{ {\cal A}^j,{\cal A}^k \}_P \bigg \}.
\eqno(12) $$

Thus the large-$N$ limit of Nahm equations is

$$ {\cal A}^j_{,t} = {1 \over 2} \epsilon^{jkl} \{ {\cal A}^k,{\cal A}^l
\}_P, \eqno(13) $$
which are precisely the large-$N$ limit of the self-duality condition. One
can see that once again SU$(\infty)$ Nahm equations (13) imply equations
(9) and (10).

\vskip 1truecm


\chapter{\bf The Moyal Deformation of Nahm Equations}

The aim of this section is to show that the reduced self-dual Yang-Mills
theory described by Nahm equations, as we saw in the above section, can be
carried over into the Weyl-Wigner-Moyal (WWM)-formalism. We will show that
the large-$N$ limit can be also achieved by taking the $\hbar \to 0$
limit.

The operator-valued equations (operator analog of Eq. (3)) for $\hat{A}^j
\in {\rm so}(3) \otimes \hat{\cal U}$ where $\hat{\cal U}$ is the Lie
algebra of anti-self-dual operators acting on the Hilbert space
$H=L^2({\bf R})$, are

$$ \hat{A}^j_{,tt} + [\hat{A}^k,[\hat{A}^k,\hat{A}^j]] = 0. \eqno(14) $$
While the quantum Gauss law is written as

$$ [\hat{A}^j_{,t},\hat{A}^j] = 0, \eqno(15) $$
where $[,]$ is the commutator.

First of all define

$$ \hat{\cal A} := i \hbar A^j. \eqno(16) $$

With Eq. (16), the Eq. (14) can be rewritten as

$$ \hat{\cal A}^j_{,tt} + {1 \over i \hbar} [\hat{\cal A}^k, {1 \over i
\hbar}[\hat{\cal A}^k,\hat{\cal A}^j]] = 0, \eqno(17) $$
and the constraint

$${1 \over i \hbar} [\hat{\cal A}^j_{,t},\hat{\cal A}^j ] = 0. \eqno(18)
$$

By simple calculations we find that equation of motion (17) can be
derived from the `quantum Lagrangian'

$$ L^{(q)} :=  Tr \bigg\{ 2 \pi \hbar \bigg[ \hat{\cal A}^j_{,t}
\hat{\cal A}^j_{,t} - {1 \over 2 \hbar^2} [\hat{\cal A}^j,\hat{\cal A}^k]
[\hat{\cal A}^j,\hat{\cal A}^k] \bigg] \bigg\} $$

$$ = 2 \pi \hbar \sum_j < \psi_j|\hat{\cal A}^j_{,t} \hat{\cal A}^j_{,t} -
{1 \over 2 \hbar^2} [\hat{\cal A}^j,\hat{\cal A}^k] [\hat{\cal A}^j,
\hat{\cal A}^k] \bigg]|\psi_j>, \eqno(19) $$
where `$Tr$' is the sum over diagonal elements with respect to an
orthonormal basis $\{ |\psi_j>\}_{j \in {\bf N}}$ in $L^2({\bf R})$ {\it
i.e.}

$$ <\psi_j| \psi_k> = \delta_{jk}, \ \ \ \ \ \sum_j |\psi_j><\psi_j| =
\hat I. \eqno(20)$$

Operator equation of motion (17) and constraint (18) are implied by  the
operator-valued Nahm equations

$$\hat{\cal A}^j_t(t)  = {1 \over 2i \hbar}
\epsilon^{jkl}[\hat{\cal A}^k(t),\hat{\cal A}^l(t)]. \eqno(21) $$

The {\it Weyl correspondence} ${\cal W}^{-1}$ establishes a one to one
correspondence between self-adjoint linear operators ${\cal B}$ acting on
Hilbert space $H= L^2({\bf R})$ and the space of real smooth functions
$C^{\infty}(\Sigma, {\bf R})$ on the phase space manifold $\Sigma$. This
correspondence $ {\cal W}^{-1}: {\cal B} \to C^{\infty}(\Sigma, {\bf R}),$
is given by

$$ {\cal A}^j(t,p,q;\hbar) \equiv {\cal W}^{-1}(\hat{\cal A}^j) :=
\int_{\infty}^{\infty} <q - {\xi \over 2}|\hat{\cal A}^j(t)|q + {\xi \over
2}> {\rm exp}\big( {i \over \hbar} \xi p \big) d\xi, \eqno(22) $$

 for all $\hat{\cal A} \in {\cal B}$ and ${\cal A} \in C^{\infty}(\Sigma,
{\bf R}).$ Define the Moyal product `$\star$' on $C^{\infty}(\Sigma, {\bf
R})$ to be $ {\cal H}_i \star {\cal H}_j := {\cal H}_i {\rm exp} ({i
{\hbar} \over 2} \buildrel{\leftrightarrow}\over{\cal P}){\cal H}_j, \ \
(i\not = j)$ where $\buildrel {\leftrightarrow}\over {\cal P} :=
{\buildrel {\leftarrow}\over{\partial} \over \partial q} {\buildrel
{\rightarrow}\over{\partial} \over \partial p} - {\buildrel
{\leftarrow}\over{\partial} \over \partial p} {\buildrel
{\rightarrow}\over{\partial} \over \partial q},$ ${\cal H}_i = {\cal H}_i
(p,q;\hbar).$

One can see easily that ${\cal W}^{-1}$ is a Lie algebra isomorphism
${\cal W}^{-1} : \big({\cal B}, [,]\big) \to \big({\cal M}, \{\cdot, \cdot
\}_M \big),$ and it is given by

$${\cal W}^{-1} \big( {1 \over i \hbar} [\hat {\cal H}_i, \hat{\cal H}_j]
\big) = {1 \over i \hbar} ({\cal H}_i \star {\cal H}_j - {\cal H}_j \star
{\cal H}_i) \equiv \{{\cal H}_i,{\cal H}_j \}_M, \eqno(23)$$ where
$\{,\}_M$ is the Moyal bracket (for details and conventions see Ref.
[10]) and ${\cal M}$ is called {\it the Moyal algebra.}

Moyal algebra ${\cal M}$ is isomorphic to a deformation of Poisson algebra
sdiff$(\Sigma)$ {\it i.e.} ${\cal M} \cong {\rm diff}_{\hbar}(\Sigma),$
where $\hbar$ is the deformation parameter. The correspondence with the
Poisson algebra is given by taking the limit $\hbar \to 0$ {\it i.e.}
$\lim_{\hbar \to 0}$ sdiff$_{\hbar}(\Sigma)=$ sdiff$(\Sigma),$ or

$$\lim_{{\hbar}\to 0} {\cal H}_i \star {\cal H}_j = {\cal H}_i {\cal H}_j
\ \ {\rm and} \
\
\lim_{{\hbar} \to 0} \{{\cal H}_i, {\cal H}_j\}_M = \{{\cal
H}_i,{\cal H}_j\}_P. \eqno(24)$$

>From the WWM-formalism an easy computation shows that operator equation of
motion (17) yields

$${\cal A}^j_{,tt} + \{{\cal A}^j,\{{\cal A}^j,{\cal A}^k\}_M\}_M = 0,
\eqno(25) $$
and the associated constraint is

$$ \{{\cal A}^j_{,t},{\cal A}^j \}_M = 0. \eqno(26) $$ Eq. (25) is the
Moyal deformation of Eq.  (3).  A computation shows that Eq. (25) can be
derived from the Lagrangian

$$ L^{(M)}_N := - \int _{\Sigma} dp dq \bigg\{ {\cal A}^j_{,t} \star {\cal
A}^j_{,t} + {1 \over 2} \{{\cal A}^j,{\cal A}^k\}_M \star \{{\cal
A}^j,{\cal
A}^K\}_M \bigg\}. \eqno(27) $$
While the Moyal deformed `conserved energy' is

$$ E^{(M)}_N = - \int_{\Sigma} dp dq \bigg\{  {\cal A}^j_{,t} \star{\cal
A}^j_{,t}  - {1 \over 2}  \{ {\cal A}^j,{\cal A}^k \}_M \star \{ {\cal
A}^j,{\cal A}^k \}_M \bigg \}. \eqno(28) $$

Using the properties (24) we find the result

$$ \lim_{\hbar \to 0} L^{(M)}_N = L^{\infty}_N, \ \ \  \lim_{\hbar \to 0}
E^{(M)}_N = E^{\infty}_N.  \eqno(29)$$
Of course we have assumed that the fields ${\cal A}^j(t,p,q,\hbar)$ can be
written as [8]

$$ {\cal A}^j(t,p,q;\hbar) = \sum_{k=0}^{\infty} \hbar^k {\cal
A}^j_{k}(t,p,q). \eqno(30)$$
and $\lim_{\hbar \to 0} {\cal A}^j(t,p,q;\hbar) = {\cal A}^j_0(t,p,q)$.

Using Weyl correspondence described as above we get a
triad of functions ${\cal A}^j = {\cal A}^j(t,p,q;\hbar) \in {\rm so}(3)
\otimes {\cal M}$ which satisfy the Moyal deformation of Nahm equations

$$ {\cal A}^j_{,t}  = {1 \over 2}
\epsilon^{jkl}\{ {\cal A}^k,{\cal A}^l\}_M. \eqno(31) $$
Using correspondence (24) we finally obtain the well known SU$(\infty)$
Nahm equations [13]

$$ {\cal A}^j_{0, t}  = {1 \over 2}
\epsilon^{jkl}\{ {\cal A}^k_0,{\cal A}^l_0\}_P, \eqno(32) $$
or after some minor manipulations

$$ {\cal A}^j_{0,t} =  \epsilon^{jkl} {\cal A}^k_{,q} {\cal A}^l_{,p}.
\eqno(33) $$

In the next section we shall intend to obtain some solutions of the above
equation (33) using WWM-formalism.

\vskip 1truecm

\subsection{Lax Pair}

Finally we find the Lax pair for the Moyal deformation of Nahm equations
(31). In order to do this, in analogy of [13], we consider the following
system of linear partial differential equations on the function $E =
E(\lambda,t,p,q) \in C^{\infty}({\bf C}^* \times T \times \Sigma)$, where
$T$ is isomorphic to the real line ${\bf R}$, ${\bf C}^* = {\bf C} -
\{0\},$ and $\Sigma \subset {\bf R}^2$

$$ i \hbar \partial_t E_{\lambda} = (- a^3 - \lambda a^-)\star
E_{\lambda}, \eqno(34a)$$

$$ i \hbar \partial_t E_{\lambda} = (a^3 + \lambda^{-1} a^+)\star
E_{\lambda}, \eqno(34b)$$ where $\lambda \in {\bf C}^*,$ $ E_{\lambda} =
E_{\lambda}(t,p,q):= E(\lambda,t,p,q)$ and $a^{\pm} := a^1 \pm i a^2$ with
$a^i = a^i(t,p,q), \ (i=1,2,3)$ independent of $\lambda.$

The integrability conditions of the above linear system read

$$ \lambda^0 \mapsto \ a^3_{,t} = {1 \over 2} \{a^+, a^-\}_M, \eqno(35a)$$

$$ \lambda^1 \mapsto \ a^-_{,t} = \{a^3, a^-\}_M, \eqno(35b)$$

$$ \lambda^{-1} \mapsto \ a^+_{,t} = \{a^+, a^3\}_M. \eqno(35c)$$

Thus one can find that Eqs. (35) written in terms of the $a^i$'s,
$i=1,2,3,$ are exactly the Moyal deformation of Nahm equations, (31).
Therefore {\it the system (34) constitutes the Lax pair of (31).}

\vskip 1truecm


\chapter{\bf Searching For  Solutions }

In this section we attempt searching for some solutions to the Moyal
deformation and SU$(\infty)$ Nahm equations (13). In order to do that we
will follow the same method which we have used to construct some solutions
to Park-Husain heavenly equations in the last reference of [10]. The
method was explained in [10,11] and it not will be repeated here.

Let $\Phi: {\cal G}_{\bf C} \to \hat{\cal I}$ be a Lie algebra
homomorphism. Here ${\cal G}_{\bf C} = {\cal G} \otimes {\bf C}$ and
$\hat{\cal I}$ is the complex Lie algebra of operators. Now we will apply
the explicit Lie algebra homomorphism $\Phi$ for the cases of ${\cal
G}_{\bf C} = {\rm su}(2)$ and ${\cal G} = {\rm sl}(2,{\bf R})$ described
in Ref. [20].

\subsection{Elliptic Curves}

The solutions of sl$(2)$-Nahm's equations (5) were obtained some years ago
[21,22,23]. Basically these solutions are solutions of the Euler equations
[23]. Some of these solutions are expressed in terms of the theory of
elliptic curves. These are

$$ A^j(t) = - \sqrt{\wp(t) - e_j}, \eqno(36)$$
where $\wp(t)$ is the Weierstrass function on the elliptic curve (with
semi-periods $\Omega_1$, $\Omega_2$ and $\Omega_3 = \Omega_1 + \Omega_2$)
determined by the two first integrals $ I_1 = (A^1)^2 -(A^2)^2 = e_2 -
e_1$ and $I_2= (A^1)^2 - (A^3)^2 = e_3 - e_1$. The fact that $A^j \in {\rm
so}(3) \otimes {\rm sl}(2)$ implies that $A^j_a$ is of the from

$$ A^j = A^j(t) = A^j_a(t) \tau_a, \ \ \ a=1,2,3 \eqno(37)$$ and should be
a diagonal matrix

$$ \pmatrix{ - \sqrt{\wp(t) -e_1} & 0 & 0 \cr
0 & - \sqrt{\wp(t) -e_2} & 0\cr
0 & 0& - \sqrt{\wp(t) - e_3}}. \eqno(38)$$

Thus solutions of Nahm equations are given by

$$ A^j(t) = \sqrt{\wp(t) - e_j} \ \tau_j. \eqno(39)$$
Applying the Lie algebra homomorphism $\Phi: {\rm su}(2) \to \hat{\cal
I}$ we get

$$\hat{\cal A}^j = \hat{\cal A}^j(t) = - i \hbar \sqrt{\wp(t) - e_j} \
\hat{\cal X}_j \in
{\rm so}(3)\otimes \hat{\cal I} \eqno(40)$$
where $\hat{\cal X}_j \equiv \Phi(\tau_j). $

Using WWM-formalism [10] we find that if Eq. (40) is solution of (17),
then

$${\cal A}^j(t,p,q;\hbar) = -\sqrt{\wp(t) - e_j} \ {\cal
X}_j(p,q;\hbar), \eqno(41)$$
where
$$ {\cal X}_j(p,q;\hbar) :=
\int_{\infty}^{\infty} <q - {\xi \over 2}|\hat{\cal X}_j|q + {\xi
\over 2}> {\rm exp}\big( {i \over \hbar} \xi p \big) d\xi, \eqno(42) $$
should be solution of the Moyal deformation of Nahm equation (31).

\subsection{The SU(2) Magnetic Monopole }

SU$(2)$ BPS monopoles can be seen to be solutions of su$(2)$ Nahm
equations and can be written as

$$ A^j = A^j(t) = A^j_a(t) \tau_a, \ \ \  a=1,2,3 \eqno(43)$$
where
$$ \tau_1 = {i \over 2} \pmatrix{ 0 & 1 \cr 1 & 0}, \ \ \  \tau_2 = {i
\over 2} \pmatrix { 0 & -i \cr i & 0}, \ \ \   \tau_3 = {i \over 2}
\pmatrix{ 1 & 0 \cr 0 & -1}, \eqno(44)$$
satisfy as usual
$$ [\tau_a,\tau_b] = \epsilon_{abc} \tau_c. $$
The matrix $A^j_a(t)$ is a diagonal matrix [21,22,23]

$$ \pmatrix{{\rm ns}(t,k) & 0 & 0 \cr
0 & {\rm ds}(t,k) & 0\cr
0 & 0& {\rm cs}(t,k)}. \eqno(45)$$
where `ns', `ds' and `cs' are elliptic functions.

Using the results of Refs. [10,20]

$$  \Phi(\tau_1) = \hat {\cal X}_1 := i \beta \hat{q} + {1\over 2{\hbar}}
(\hat{q}^2 - 1)\hat{p}, \eqno(46a)$$

$$  \Phi(\tau_2) = \hat {\cal X}_2 := - \beta \hat{q} + {i\over 2{\hbar}}
(\hat{q}^2 + 1)\hat{p}, \eqno(46b)$$

$$ \Phi(\tau_3) = \hat {\cal X}_3 := - i \beta \hat{1} - {1\over {\hbar}}
\hat{q} \hat{p}, \eqno(46c)$$
where $\beta \in {\bf R}$ is any constant; $\hat{q}$ and $\hat{p}$ are the
position and momentum operators, respectively. Then, the complex functions
${\cal A}^j = {\cal A}^j(t,p,q)$ should be a solution of the Moyal
deformation of the Nahm equations.

$${\cal A}^1(t,p,q;\hbar) = {\rm ns}(t,k)[{i \over 2}p(q^2 - 1) -
\hbar q (\beta + {1\over 2})], \eqno(47a)$$

$${\cal A}^2(t,p,q;\hbar) = {\rm ds}(t,k)[-{1 \over 2}p(q^2 + 1) -
i\hbar q (\beta + {1\over 2})], \eqno(47b)$$

$${\cal A}^3(t,p,q;\hbar) = {\rm cs}(t,k)[-ipq + \hbar(\beta + {1\over
2})]. \eqno(47c)$$

Taking the $\hbar \to 0$ limit in the above equations we have

$${\cal A}^1_0(t,p,q;\hbar) = {\cal L}_1(p,q){\rm ns}(t,k), \eqno(48a)$$

$${\cal A}^2_0(t,p,q;\hbar) = {\cal L}_2(p,q){\rm ds}(t,k), \eqno(48b)$$

$${\cal A}^3_0(t,p,q;\hbar) = {\cal L}_3(p,q){\rm cs}(t,k), \eqno(48c)$$
where ${\cal L}_1 = {i \over 2}p(q^2 - 1)$, ${\cal L}_2 = -{1 \over
2}p(q^2 + 1)$ and ${\cal L}_3 =  -ipq$. It is immediate to see that
the functions ${\cal L}_j$ satisfy the algebra

$$ \{{\cal L}_i,{\cal L}_j \}_P = \epsilon_{ijk}{\cal L}_k. $$
It is very interesting to see that Eqs. (48) are precisely the solutions
obtained recently by Hashimoto et al. [24] as a hyper-K\"ahler metric on
$\Sigma \times {\bf R}^2$ which is a particular solution of the
Ashtekar-Jacobson-Smolin equations! [7]. Following the philosophy of Ref.
[11], the above solution might be called {\it gravitational monopole.}
Thus we have found a case where a solution of BPS monopole Nahm equations
corresponds to a self-dual metric.  This correspondence deserves a more
careful study and we leave it for a separated paper.

\subsection{ The $SL(2;{\bf R})$ Magnetic Monopole}

A similar situation as the above results occurs for the SL$(2;{\bf R})$
case. Now the solution of sl$(2,{\bf R})$ Nahm's equations (5) takes the
form

$$ A^j = A^j(t) = A^j_a(t) \tau_a, \ \ \ \ \  a=1,2,3,$$
where
$$ \tau_1 = {1 \over 2} \pmatrix{ 0 & 1\cr 1 & 0}, \ \ \  \  \tau_2 = {1
\over 2} \pmatrix{ 0 & -1\cr 1 & 0}, \ \ \ \   \tau_3 = {1 \over 2}
\pmatrix{ 1 & 0\cr 0 & -1}, \eqno(49)$$
satisfy
$$ [\tau_1, \tau_2] = \tau_3, \ \ \ \ [\tau_2, \tau_3] = \tau_1, \ \ \ \
[\tau_3, \tau_1] = - \tau_2.$$

Following Ward's paper [21], the solutions of sl$(2;{\bf R})$ Nahm
equation are

$$A^1 = k {\rm sn}(t,k), \ \ \  \ A^2 = k {\rm cn}(t,k), \ \ \ \  A^3 =
{\rm dn}(t,k), \eqno(50) $$
where `sn', `cn' and `dn' are elliptic functions and $k$ is a constant.

>From Refs. [20,10] we define the Lie algebra homomorphism $ \Phi:  {\rm
sl}(2;{\bf R}) \to \hat{\cal U}$ by

$$ \Phi(\tau_1) = \hat{\cal X}_1 := {i \over 4} \big( {\hat{p}^2 \over
{\hbar}^2} + {\delta \over \hat{q}^2} - \hat{q}^2\big), \eqno(51a)$$

$$ \Phi(\tau_2) = \hat{\cal X}_2 := {i \over 4} \big( {\hat{p}^2 \over
{\hbar}^2} + {\delta \over \hat{q}^2} + \hat{q}^2\big), \eqno(51b)$$

$$ \Phi(\tau_3) = \hat{\cal X}_3 := {i \over 2} \big( {\hat{q} \hat{p}
\over
{\hbar}} - {i \over 2}\big), \eqno(51c)$$
where $\delta \in \Re$ is a constant.

Defining the functions $ {\cal X}_a = {\cal X}_a(p,q)$ according to (42)
and proceeding as above we get the real solution of the Moyal deformation
of Nahm equation to be

$${\cal A}^1(t,p,q;\hbar) = k {\rm sn}(t,k)[- {\hbar^{-1} \over 4}p^2 -
{\hbar \delta \over 4} q^{-2} + {\hbar \over 4} q^2], \eqno(52a)$$

$${\cal A}^2(t,p,q;\hbar) = k {\rm cn}(t,k)[- {\hbar^{-1} \over 4}p^2 -
{\hbar \delta \over 4} q^{-2} - {\hbar \over 4} q^2], \eqno(52b)$$

$${\cal A}^3(t,p,q;\hbar) = {\rm dn}(t,k)[-{1 \over 2}pq ]. \eqno(52c)$$

One can see that once again this sl$(2;{\bf R})$ BPS monopole defines the
Moyal deformation but in this case the solution is not analytic in
$\hbar$. Thus, in this case doesn't exist an associated self-dual metric.

\vskip 1truecm


\chapter{\bf Final Remarks}

In this paper we have found another example, different from self-dual
gravity, where the WWM-formalism gets some solutions for large-$N$
physical systems. In particular we have defined the Moyal deformation of
Nahm equations and we have found the corresponding SU$(\infty)$ Nahm
equations as the $\hbar \to 0$ limit instead of the usual one $N \to
\infty$. After that we found some solutions of SU$(\infty)$ Nahm equations
via the WWM-formalism. We have found strong evidence of the existence of
the correspondence between the SU$(2)$ BPS magnetic monopoles which are
solutions of Nahm equations in agreement to [2] and hyper-K\"ahler metrics
on $\Sigma \times {\bf R}^2$. We leave this important issue for a
forthcoming paper. We think that Strachan geometry of multidimensional
geometry [25] might be important to consider the whole and global issue
where topological field theories and cobordism structure [26] are very
important.

\vskip 1truecm

\noindent
{\it Remark}

Some recent works on deformations of Nahm equations were given by Hoppe
[27] and by Castro [28,29]. Hoppe shown that the existence of infinitely
constants of motion is preserved under non-local deformations of some
surfaces generated by time harmonic motions. In Ref. [28] Castro uses the
WWM-formalism to construct Moyal deformations of the self-dual membrane in
terms of the 3D Moyal deformation of the continuous Toda theory. Finally
in Ref. [29] it was obtained the $q$-Moyal deformation of the self-dual
membrane and very interesting explicit solutions were found. Embedding of
Moyal-Toda chain into the SU$(\infty)$ Moyal-Nahm equation was given as
well. We are grateful to the referee for pointing out the papers [27-29].

\vskip 1truecm

\leftline{\ilis Acknowledgements}

We would like to thank M. Przanowski and J. Tosiek for several useful
discussions and suggestions and Carlos Castro for useful communications.
The work of H. G-C. is supported by a Postdoctoral CONACyT fellowship in
the program {\it Programa de Posdoctorantes: Estancias Posdoctorales en
el Extranjero para Graduados en Instituciones Nacionales: 1996-1997} and
in part by the Academia Mexicana de Ciencias (AMC) in the program {\it
Estancias de Verano para Investigadores J\'ovenes.} The work of J. P. is
supported in part by a CONACyT grant.

\vskip 2truecm
\centerline{\Hugo References}

\item{[1]} N. Seiberg and E. Witten, Nucl. Phys. {\bf B426} (1994) 19,
hep-th/9407087; Nucl. Phys. {\bf B431} (1994) 484, hep-th/9408099.

\item{[2]} R.S. Ward and R.O. Wells Jr, {\it Twistor Geometry and Field
Theory}, Cambridge University Press, Cambridge (1990); E. Corrigan and P.
Goddard, Commun. Math. Phys. {\bf 80} (1981) 575; N.J. Hitchin, Commun.
Math. Phys. {\bf 83} (1982) 579, {\bf 89} (1983) 145; S.K. Donaldson,
Commun. Math. Phys. {\bf 96} (1985) 387; M.F. Atiyah and N.J. Hitchin,
{\it The Geometry and Dynamics of Magnetic Monopoles}, Princeton
University Press, Princeton (1988); H. Nakajima, ``Monopoles and Nahm
Equations'', {\it Einstein Metrics and Yang-Mills Connections} (Sanda
1990), 193, Lecture Notes in Pure and Applied Mathematics {\bf 145},
Dekker, New York (1993).

\item{[3]} W. Nahm, Phys. Lett. {\bf 90B} (1980) 413; ``The Construction
of All Self-dual Multimonopoles by the ADHM Method'' in {\it Monopoles in
Quantum Field Theory}, Eds. N. Craigie et al. World Scientific, Singapore
(1982); ``Mathematical Structures Underlying Monopoles in Gauge
Theories'', Ed. N. Craigie, World Scientific, Singapore (1986).

\item{[4]} D-E Diaconescu, ``D-Branes, Monopoles and Nahm Equations'',
hep-th/608163.

\item{[5]} J-S. Park, ``Monads and D-instantons'', hep-th/9612096.

\item{[6]} A. Hanany and E. Witten, ``Type IIB Superstrings, BPS
Monopoles, And Three-Dimensional Gauge Dynamics'', IASSNS-HEP-96/121,
hep-th/9611230.

\item{[7]} A. Ashtekar, T. Jacobson, L. Smolin, Commun. Math. Phys. {\bf
115} (1988) 631-648;L.J. Mason and E.T. Newman, Commun. Math. Phys. {\bf
121} (1989) 659.

\item{[8]} I.A.B. Strachan, Phys. Lett. B {\bf 283} (1992) 63.

\item{[9]} K. Takasaki, J. Geom. Phys. {\bf 14} (1994) 111; {\bf 14}
(1994) 332.

\item{[10]} J.F. Pleba\'nski, M. Przanowski, B. Rajca and J. Tosiek, Acta
Phys. Pol. B {\bf 26} (1995) 889; J.F. Pleba\'nski, M.  Przanowski and J.
Tosiek, Acta Phys. Pol. B {\bf 27} (1996) 1961; J.F.  Pleba\'nski and M.
Przanowski, ``The Universal Covering of Heavenly Equations Via
Weyl-Wigner-Moyal Formalism'' in {\it Gravitation, Electromagnetism and
Geometrical Structures} For the 80th birthday of A. Lichnerowitz, Ed.
Giorgio Ferrarese, Pitagora Editrice, Bologna (1996); ``The Lagrangian of
a self-dual gravitational field as a limit of the SDYM Lagrangian'', Phys.
Lett. A {\bf 212} (1996) 22, hep-th/9605233; J.F. Pleba\'nski M.
Przanowski and H.  Garc\'{\i}a-Compe\'an, ``From Principal Chiral Model to
Self-dual Gravity'', Mod. Phys. Lett. {\bf A11} (1996) 663,
hep-th/9509092.

\item{[11]} H. Garc\'{\i}a-Compe\'an, J. F. Pleba\'nski and M.
Przanowski,  ``Further Remarks on the Chiral Model Approach to Self-dual
Gravity'', Phys. Lett. A {\bf 219} (1996) 249, hep-th/9512013.

\item{[12]} K. Uhlenbeck, J. Diff. Geom. {\bf 30} (1989) 1.

\item{[13]} R.S. Ward, Phys. Lett. {\bf B 234} (1990) 81; J. Geom. Phys.
{\bf 8} (1992) 317.

\item{[14]} E. Witten, J. Geom. Phys. {\bf 8} (1992) 327.

\item{[15]} E.G. Floratos and G.K. Leontaris, Phys. Lett. B  {\bf 223}
(1989) 153.

\item{[16]} E.G. Floratos, Phys. Lett. B {\bf 228} (1989) 335.

\item{[17]} D.B. Fairlie, P.. Fletcher and C.K. Zachos, J. Math. Phys. 31
(1990) 1088.

\item{[18]} E.G. Floratos, J. Iliopoulos and G. Tiktopoulos, Phys. Lett.
{\bf B217} (1989) 285.

\item{[19]} J. Hoppe, M.I.T. Ph. D. Thesis (1982); Constraints Theory and
Relativistic Dynamics- Florence 1986, G. Longhi and L. Lusanna eds. p.
267, World Scientific, Singapore (1987); Phys. Lett. B {\bf 215} (1988)
706.

\item{[20]} K.B. Wolf, ``Integral Transform Representations of
SL$(2,\Re)$'', in {\it Group Theoretical Methods in Physics}, Cocoyoc
M\'exico 1980, ed. K.B. Wolf (Springer-Verlag, 1980) pp. 526-531.

\item{[21]} R.S. Ward, J. Phys. A {\bf 20} (1987) 2679.

\item{[22]} S. Chakravarty, M.J. Ablowitz and P.A. Clarkson, Phys. Rev.
Lett. {\bf 65} (1990) 1085.

\item{[23]} F. Guil and M. Ma\~nas, ``Nahm Equations and Self-dual
Yang-Mills Equations'', Phys. Lett. B {\bf 302} (1993) 431-434.

\item{[24]} Y. Hashimoto, Y. Yasui, S. Miyagi and T. Otsuka,
``Applications of the Ashtekar Gravity to Four Dimensional Hyper-K\"ahler
Geometry and Yang-Mills Instantons'', hep-th/9610069.

\item{[25]} I.A.B. Strachan, ``A Geometry for Multidimensional
Integrable Systems'', J. Geom. Phys. (1996), to appear, hep-th/9604142.

\item{[26]} S.K. Donaldson, ``Complex Cobordism, Ashtekar's Equations and
Diffeomorphisms'' in {\it Symplectic Geometry}, ed. D. Salamon, London
Math. Soc. (1992) 45-55.

\item{[27]} J. Hoppe, ``On the Deformation of Time Harmonic Flows'',
hep-th/9612024.

\item{[28]} C. Castro, ``A Moyal Quantization of the Continuous Toda
Field'', hep-th/9703094.

\item{[29]} C. Castro, ``SU$(\infty)$ $q$-Moyal-Nahm Equations and Quantum
Deformations of the Self-Dual Membrane'', hep-th/9704031.

\endpage
\end